\documentclass[aps,prl,twocolumn,showpacs,superscriptaddress]{revtex4}

\usepackage{graphicx}

\usepackage{graphicx}
\usepackage{color}

\begin{document}

%\preprint{}

\title{Surface structure of Bi$_{2}$Se$_{3}$(111) determined by low-energy electron diffraction and surface X-ray diffraction}


\author{Diogo Duarte dos Reis}
\affiliation{Departamento de F\'isica, ICEx, Universidade Federal de Minas Gerais, 30123-970 Belo Horizonte, MG, Brazil}
\author{Lucas Barreto}
\author{Marco Bianchi}
\affiliation{Department of Physics and Astronomy, Interdisciplinary Nanoscience Center, Aarhus University, 8000 Aarhus C, Denmark}
\author{Guilherme Almeida Silva Ribeiro}
\author{Edmar Avellar Soares}
\author{Wendell Sim\~oes e Silva}
\author{Vagner Eust\'aquio de Carvalho}
\affiliation{Departamento de F\'isica, ICEx, Universidade Federal de Minas Gerais, 30123-970  Belo Horizonte, MG, Brazil}
\author{Jonathan Rawle}
\author{Moritz Hoesch}
\author{Chris Nicklin}
\affiliation{Diamond Light Source Ltd, HSIC, Didcot, Oxfordshire, OX11 0DE, United Kingdom}
\author{Willians Principe Fernandes}
\affiliation{Departamento de Ci\^encias Naturais, Universidade Federal de S\~ao Jo\~ao Del-Rei, MG, Brazil}

\author{Jianli Mi}
\author{Bo Brummerstedt Iversen}
\affiliation{Center for Materials Crystallography, Department of Chemistry, Interdisciplinary Nanoscience Center, Aarhus University,
8000 Aarhus C, Denmark}

\author{Philip Hofmann}
\affiliation{Department of Physics and Astronomy, Interdisciplinary Nanoscience Center, Aarhus University, 8000 Aarhus C, Denmark}


\date{\today}

\begin{abstract}
The surface structure of the prototypical topological insulator Bi$_2$Se$_3$ is determined by low-energy electron diffraction and surface X-ray diffraction at room temperature. Both approaches show that the crystal is terminated by an intact quintuple layer. Specifically, an alternative termination by a bismuth bilayer is ruled out. Surface relaxations obtained by both techniques are in good agreement with each other and found to be small. This includes the relaxation of the van der Waals gap between the first two quintuple layers.
\end{abstract}

%\keywords{}

\maketitle

%\section{Introduction}
%\label{intro}
Bismuth selenide and bismuth telluride have recently attracted considerable attention as prototypical topological insulators. The electronic band structure has a negative band gap at $\Gamma$, resulting in an odd number of closed surface state Fermi contours around the centre of the surface Brillouin zone, i.e. in a particularly simple manifestation of topologically protected surface states \cite{Noh:2008,Zhang:2009,Xia:2009,Hsieh:2009c}.

\begin{figure}[h!]
\includegraphics[width=.45\textwidth]{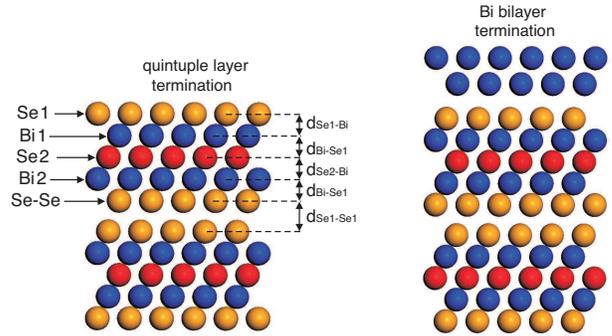}%
\caption{Left: Side view of a bulk-terminated Bi$_2$Se$_3$(111) crystal with indication of the terminations tested and the notation for the interlayer distances. Right: One possible bilayer-terminated surface (different stacking possibilities are not shown). \label{fig1}}
\end{figure}

Bi$_2$Se$_3$ has a layered crystal structure, made up from Se-Bi-Se-Bi-Se quintuple layers (QLs), separated by van der Waals gaps (see Figure \ref{fig1}). The (111) surface of the material is the surface parallel to these QLs and can be prepared easily by cleaving the crystal with scotch tape. It therefore appears likely that this cleaving process takes place between two QLs and the surface is thus terminated by an intact QL.
However, this termination has recently been questioned by a low energy ion scattering investigation which revealed that a surface obtained by cleaving at room temperature, or left for some time at low temperature, develops a strong enrichment of Bi, consistent with a termination by a bismuth bilayer on top of the last QL of Bi$_2$Se$_3$  \cite{He:2013}. Such a change of the surface termination (of Bi$_2$Se$_3$ or Bi$_2$Te$_3$)  should lead to a drastic modification of the surface electronic structure  \cite{Hirahara:2011,He:2013,Miao:2013} compared to the single Dirac cone usually observed \cite{,Xia:2009,Hsieh:2009c}. Angle-resolved photoemission (ARPES) spectra for Bi$_2$Se$_3$ cleaved at room temperature do not show this more complicated electronic structure \cite{Hatch:2011}, but it can be created on both Bi$_2$Te$_3$ and Bi$_2$Se$_3$ by depositing a bilayer of bismuth on purpose \cite{Hirahara:2011,Miao:2013}. 

Even for the bulk terminated by a QL, details of the structural relaxations are crucial for the electronic structure. It has been observed early  that ARPES spectra from Bi$_2$Se$_3$ change with time after cleave \cite{Hsieh:2009c}. The change manifests itself as a shift of all the bands to higher binding energy and the appearance of new two-dimensional states on the surface \cite{Bianchi:2010b,Bianchi:2011}. An initial interpretation of this was a structural relaxation of the van der Waals gaps below the surface \cite{Noh:2008,Hsieh:2009c} and it was shown theoretically that an increased van der Waals gaps could indeed give rise to two-dimensional electronic states that are similar to those observed by ARPES \cite{Menshchikova:2011,Vergniory:2012}. In related layered systems with van der Waals gaps, such a surface relaxation can in fact reproduce observed splittings of ARPES band dispersions \cite{Hoesch:2009}. Intercalating of atoms into Bi$_2$Se$_3$ to increase the van der Waals gap spacing on purpose, however, did not lead to changes in the electronic structure \cite{Bianchi:2012b}, and an alternative interpretation of the phenomenon is the formation of two-dimensional electron gases near the surface caused by an adsorbate-induced band bending \cite{Bianchi:2010b,Bianchi:2011,Bahramy:2012,King:2011}.

A detailed structural determination of the Bi$_2$Se$_3$(111) surface cleaved at room temperature is therefore called for and presented here. We use two complementary and powerful structural techniques, low-energy electron diffraction (LEED) and surface X-ray diffraction (SXRD).

The Bi$_2$Se$_3$ crystals were grown by standard methods described elsewhere \cite{Bianchi:2010b}. The bulk structure was determined by X-ray diffraction at room temperature. To this end, a fine powder was filed from the crystal rod and diffraction experiments were performed on a STOE powder diffractometer using Cu K$_{\alpha1}$ radiation in transmission geometry. The bulk structure parameters were analysed by Rietveld refinement and found to be in good agreement with the literature \cite{Nakajima:1963}. 
These structural parameters were used as starting and reference points for the surface structure determination. 
LEED and SXRD experiments were performed in ultra-high vacuum (UHV) chambers with a base pressure of $\approx  10^{-10}$ Torr.  The samples were cleaved at room temperature in a loadlock with a somewhat inferior pressure. X-ray photoemission spectroscopy performed in the LEED chamber did not show any detectable contaminations. SXRD data were taken at  beamline I07 of the Diamond Light Source, using 20~keV X-rays and a UHV chamber mounted on a large '2+3' diffractometer \cite{Vlieg:1998}. Scattered X-rays were collected using a two-dimensional detector (Pilatus) enabling fast data acquisition. The specular reflectivity (00 rod) was collected using a conventional $\Theta-2\Theta$ scan, whilst all non-specular data was recorded using a fixed X-ray incidence angle of 1$^{\circ}$. Note that data were acquired over a time span in the order of hours, whereas Ref.  \cite{He:2013} reports an increased Bi concentration near the surface, interpreted as a bilayer formation, immediately after cleaving the sample at room temperature.  
%
%\begin{figure}[h!]
%\includegraphics[width=8cm]{leedpattern.jpg}%
%\caption{Bi$_{2}$Se$_{3}$ LEED pattern at 161 eV.\label{leedpat}}
%\end{figure}

Full dynamic LEED \textit{I(V)} model calculations were performed using a modified version of the Symmetrised Automated Tensor LEED (SATLEED) computer package \cite{michel1,VanHove:1986}. The potential and the electron scattering phase-shifts for the Bi$_{2}$Se$_{3}$ (111) surface were calculated using the optimised muffin-tin potential method \cite{Rundgren:2003}, an approach recently used to successfully determine complex metal oxide surfaces \cite{Nascimento:2007,Pentcheva:2008,Nascimento:2009}. Specific phase-shift sets were calculated for selenium and for bismuth atoms, depending on their surface and bulk positions. The  \textit{I(V)} model calculations converged when using 12 phase shifts ($l_{max}=11$); and 13 phase shifts were used in the final calculations. Convergence for a lower number of phase shifts than in e.g. Ref.  \cite{Fukui:2012} is probably caused by two factors: one is the different method of phase shift calculations and the other is the lower maximum electron energy in the experiment, which was found sufficient due to the higher temperature. Debye temperatures were obtained from Ref. \cite{Shoemake:1969}. The real and imaginary parts of the optical potential were set to $V_{0} = 10.0$~eV and $V_{0i}=$-5.0~eV, respectively.

SXRD crystal truncation rod intensities were extracted by numerically integrating the background-subtracted spot in a well defined region of interest on the detector image. The structure factors were calculated by applying correction factors to account for the polarisation of the X-ray beam, the rod interception, and the area of the sample contributing to the scattered intensity, and then taking the square root of the corrected intensity \cite{Schleputz:2011}. Subsequent analysis of the data was undertaken using the ANAROD code \cite{Vlieg:2000}.

\begin{figure}[h!]
\includegraphics[width=.5\textwidth]{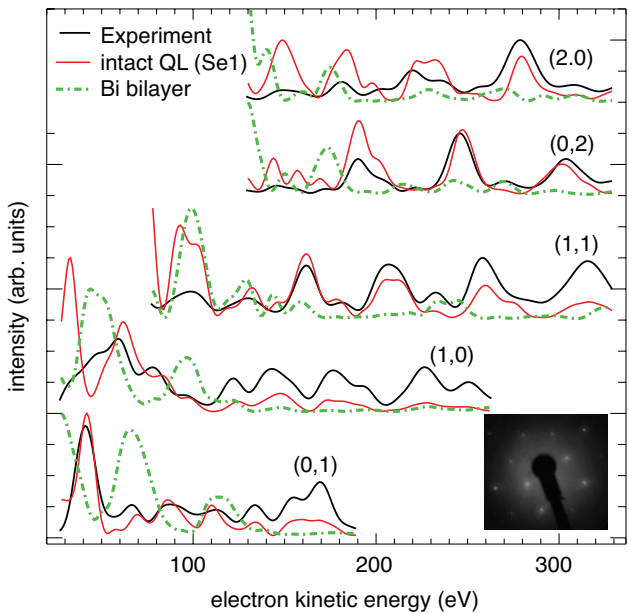}%
\caption{LEED experimental and theoretical \textit{I(V)} curves for the best structural model.  The inset shows the LEED pattern at 161 eV. \label{fig2}}
\end{figure}

\begin{figure}[h!]
\includegraphics[width=.45\textwidth]{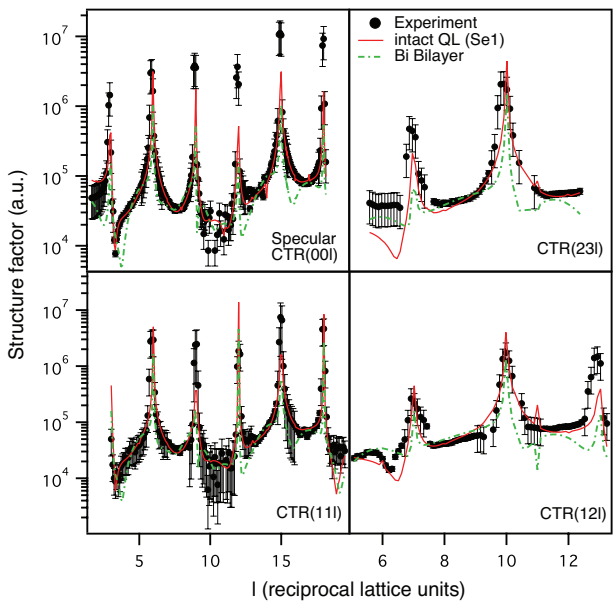}%
\caption{SXRD data from four different crystal truncation rods (CTRs) and fit to two different terminations.  \label{fig3}}
\end{figure}

The structure determination was performed by a quantitative comparison between the experimental and theoretical \textit{I(V)} curves and crystal truncation rod intensities. The agreement between experimental data and model calculations was quantified using the Pendry reliability factor ($R_{P}$) \cite{Pendry:1980} and by $\chi^2$ for LEED and SXRD, respectively. In the first step of the structure analysis, five bulk truncated trial models were tried: the  different possible possible bulk terminations (\textit{Se1, Bi1, Se2, Bi2,} and \textit{Se-Se}, see Fig. \ref{fig1}) as well a bismuth bilayer atop of the truncated bulk crystal. The resulting agreement is shown Table \ref{rptable1}. After that, an optimisation procedure was used to adjust the structural parameters as well as the inner potential (for LEED) to find the optimum structure, namely the one that leads to the lowest $R_{p}$ or $\chi^2$. For the LEED analysis of the \textit{Se1} structure, the surface Debye temperatures of the first four layers were refined, too, but this gave only rise to a very small change of $R_P$. The final values of  $R_{p}$ and $\chi^2$ for all structural models are also listed in Table \ref{rptable1}. Regarding the Bi bilayer atop model \cite{He:2013}, all the nine stacking possibilities were tested and the structural parameters of the bilayer were refined. The best values of $R_{P}$ and $\chi^2$ are given in Table \ref{rptable1}. The final comparison between measured and calculated diffraction data are shown in Figures \ref{fig2} and \ref{fig3}, both for the best fit obtained with the  intact QL (\textit{Se1}) termination as well as for the best agreement that can be reached with an optimised Bi bilayer termination.

\begin{table}%[H] add [H] placement to break table across pages
\caption{LEED Pendry R-factor and SXRD $\chi^2$ for different trial models and for non-optimised (truncated bulk) as well as optimised parameters.  \label{rptable1}}
\begin{ruledtabular}
\begin{tabular}{c c c c c}
      & \multicolumn{2}{c}{LEED $R_{P}$} & \multicolumn{2}{c}{SXRD $\chi^2$} \\
\hline      
Model & Bulk & Optimised & Bulk & Optimised \\
\hline
\textit{Se1}        & 0.74 & \textbf{0.25}  &3.15  & \textbf{2.44} \\
\textit{Bi1}        & 0.83      & 0.61 & 9.50 & 4.68 \\
\textit{Se2}        & 0.91      & 0.70 & 30.32 & 10.20 \\
\textit{Bi2}        & 0.80      & 0.65 & 30.31 & 7.02 \\
\textit{Se-Se}      & 0.98      & 0.80 & 9.46 & 7.05 \\
\textit{Bi atop \cite{He:2013}} & 1.00 & 0.84 & 11.88 & 6.31 \\
\end{tabular}
\end{ruledtabular}
\end{table}

For both LEED and SXRD, the termination with an intact QL, i.e. the \textit{Se1} model, gives the best agreement between experimental and simulated diffraction data. The detailed structural relaxations for this termination are given in Table \ref{leedatomic}.  The agreement between the two structural techniques is excellent, apart from the third interlayer distance $d_{Se2-Bi}$ where LEED finds a slightly larger expansion than SXRD, and the fourth interlayer distance $d_{Bi-Se1}$   where LEED points towards a small contraction of the interlayer spacing whereas SXRD shows an expansion. Both techniques are consistent in finding a small contraction in the first interlayer spacing $d_{Se1-Bi}$ and a small expansion of the van der Waals gap distance $d_{Se1-Se1}$.

\begin{table}%[H] add [H] placement to break table across pages
\caption{Interlayer distances found from LEED  and SXRD (see Fig. \ref{fig1}) and the corresponding bulk distances. All values are in given in \AA ngstroms. The uncertainties for the bulk values are smaller than 0.01 \AA. \label{leedatomic}}
\begin{ruledtabular}
\begin{tabular}{c c c c}
Interlayer Distances& Bulk Value & LEED & SXRD \\%& DFT \\
\hline
$d_{Se1-Bi}$  & 1.62 & 1.56 $\pm$ 0.03 & 1.51 $\pm$ 0.05 \\%1.58 & \\
$d_{Bi-Se2}$  & 1.95 & 1.96 $\pm$ 0.03 & 1.94 $\pm$ 0.06 \\%1.93 & \\
$d_{Se2-Bi}$  & 1.95 & 2.01 $\pm$ 0.04 & 1.91 $\pm$ 0.05 \\%1.92 & \\
$d_{Bi-Se1}$  & 1.62 & 1.53 $\pm$ 0.05 & 1.72 $\pm$ 0.04 \\%1.61 & \\
$d_{Se1-Se1}$ & 2.42 & 2.51 $\pm$ 0.08 & 2.50 $\pm$ 0.06 \\%2.31 & \\
\end{tabular}
\end{ruledtabular}
\end{table}

From the results of this structural determination, we can draw the following conclusions: Both techniques unambiguously show that, at room temperature, the Bi$_2$Se$_3$(111) surface is terminated with an intact QL rather than covered by a Bi bilayer. An intact QL termination also suggest that this termination prevails at lower temperature, since less thermal activation energy is available for a major structural rearrangement. The observed termination is at variance with the recent low-energy ion scattering result report in Ref. \cite{He:2013} but consistent with other experimental evidence. Notably, the surface electronic structure for a Bi bilayer on Bi$_2$Se$_3$(111) is quite different \cite{Miao:2013} from the simple single Dirac cone observed for the surface terminated by a van der Waals gap. While only low-temperature ARPES data for an on-purpose prepared bilayer of Bi on Bi$_2$Se$_3$(111) are available, it is unlikely that a room temperature measurement such as in Ref. \cite{Hatch:2011} would miss the electronic structure change and the additional bands caused by the bilayer. Finally, it is interesting to note that a recent room temperature SXRD structural determination of thin Bi$_2$Te$_3$ films on Si has a also shown the films to have a intact QL termination towards vacuum \cite{Liu:2013b}.

The details of the surface relaxations are also interesting. The contraction of the first interlayer spacing $d_{Se1-Bi}$ is of the order of 5\% and such a large contraction is unusual for a closed packed surfaces. On closed-packed simple metal surfaces, small expansions of the first interlayer spacing are more frequently found than contractions \cite{Hofmann:1996e}. On Bi$_2$Te$_3$(111), a small contraction of 1 \% was reported \cite{Fukui:2012}. More importantly, neither LEED nor SXRD find a significant expansion on the first van der Waals gap spacing. Both point towards a small expansion in the order of 4 \%, far smaller than the expansions of 20 - 40 \% required to explain the observed two-dimensional electronic states in the conduction band as caused by an interlayer expansion \cite{Menshchikova:2011,Vergniory:2012}. This can be taken as additional evidence that band bending, an not surface relaxation, is the cause for the appearance of new two-dimensional electronic states near the surface of Bi$_2$Se$_3$ \cite{Bianchi:2010b,Bianchi:2011,Bahramy:2012,King:2011}.

In summary, LEED and the SXRD reveal that the room temperature structure of  Bi$_2$Se$_3$(111) is that of a surface terminated by an intact QL, i. e. as expected for a surface cleaved in a van der Waals gap. Both techniques agree on a small expansion of the first van der Waals gap below the surface but the size of the expansion is far smaller than what would be required to bring about a dramatic change of the surface electronic structure.

%
%
%\begin{figure}[h!]
%\includegraphics[width=12cm]{bi2se3_gapinvestigation.jpg}%
%\caption{Van der Waals gap investigation by LEED.\label{leedgap}}
%\end{figure}
\begin{acknowledgments}
We gratefully acknowledge financial support by the VILLUM foundation, the Danish National Research Foundation, the Carslberg foundation, 
CNPq, CAPES, FAPEMIG, and the Laborat\'orio Nacional de Luz S\'incrotron (LNLS), as well as travel support from the Diamond Light Source Ltd. under proposal SI7522. \end{acknowledgments}

%\bibliography{groupreferences_new,local}
\providecommand{\noopsort}[1]{}\providecommand{\singleletter}[1]{#1}%

\end{document}